\documentclass{elsart}
\usepackage[english]{babel}
\usepackage[latin1]{inputenc}
\usepackage{latexsym}
\usepackage{amssymb}
\usepackage{amsmath}
\usepackage{rotating,graphicx}
\usepackage{epsfig}
\usepackage{psfrag}
\usepackage{caption}
\usepackage{everysel}
\usepackage{keyval}
\usepackage{ragged2e}

\usepackage{caption}

\newcommand{\ROOT}{\textsf{ROOT }}
\newcommand{\MAT}{\textsl{MATHEMATICA}$^\circledR$}

\newcommand{\beq}[0]{\begin{equation}}
\newcommand{\eeq}[0]{\end{equation}}

\begin{document}
\newlength{\caheight}
\setlength{\caheight}{9pt}
\multiply\caheight by 7
\newlength{\secondpar}
\setlength{\secondpar}{\hsize}
\divide\secondpar by 3
\newlength{\firstpar}
\setlength{\firstpar}{\secondpar}
\multiply\firstpar by 2

\parbox[0pt][\caheight][t]{\firstpar}{
  {\small \shortstack[l]{
     Corresponding author: Enrica Cisana\\
     e-mail: Enrica.Cisana@pv.infn.it
  }}
}
\hfill
\parbox[0pt][\caheight][t]{\secondpar}{
  \rightline
      {\shortstack[l]{
	  FNT/T 2007/05}}
}
\begin{frontmatter}
\title{A Comparative Study of \\ Stochastic Volatility Models}
\author[1,2]{Enrica Cisana}
\author[1]{Lorenzo Fermi}
\author[1,2,3]{Guido Montagna}
\author[2,3]{Oreste Nicrosini}
\address[1]{Dipartimento di Fisica Nucleare e Teorica, Universit\`a di Pavia\\
Via A. Bassi 6, 27100, Pavia, Italy}
\address[2]{Istituto Nazionale di Fisica Nucleare, sezione di Pavia\\
Via A. Bassi 6, 27100, Pavia, Italy}
\address[3]{Istituto Universitario di Studi Superiori (IUSS)\\
Viale Lungo Ticino Sforza 56, 27100, Pavia, Italy}

\begin{abstract}
The correlated stochastic volatility models constitute a natural extension of the Black and 
Scholes-Merton framework: here the volatility is not a constant, but a stochastic process correlated 
with the price log-return one. At present, several stochastic volatility models are discussed in the 
literature, differing in the dynamics attached to the volatility. The aim of the present work is to 
compare the most recent results about three popular models: the Vasicek, Heston and exponential 
Ornstein-Uhlenbeck models. We analyzed for each of them the theoretical results known in the literature 
(volatility and return distribution, higher-order moments and different-time correlations) in order 
to test their predictive effectiveness on the outcomes of original numerical simulations, paying 
particular attention to their ability to reproduce empirical statistical properties of prices. 
The numerical results demonstrate that these models can be implemented maintaining all their 
features, especially in view of financial applications like market risk management or option 
pricing. In order to critically compare the models, we also perform an empirical analysis of 
financial time series from the Italian stock market, showing the exponential Ornstein-Uhlenbeck 
model's ability to capture the stylized facts of volatility and log-return probability distributions.
\end{abstract}


\end{frontmatter}

\newpage
\section{Introduction}
\label{s:intro}

It is well documented by many statistical studies~\cite{mantegna,bouchaud} that
financial markets exhibit a very complex dynamics. In parallel
with this empirical research, various theoretical models, based on non-Gaussian dynamics for the 
time evolution of price returns, have been proposed in the literature to cope with the non-trivial
stylized facts observed in real markets. Among them, the stochastic volatility models (SV)~\cite{fouque,scott,stein,heston} 
have receivedparticular attention because of their analytical tractability and
parsimonious use of free parameters. Since much work has
been devoted to the derivation of (semi)-analytical results,
the aim of the present paper is mainly to test and critically
compare the most recent theoretical results by means of detailed
numerical simulations, as well as to provide a further contribution
to the empirical analysis of SV models already existing in the
literature.
\section{Theoretical approach: the models}
\label{s:theoretical}

The idea behind SV models is that the volatility $\sigma$, a constant in the Black and Scholes-Merton framework, is 
itself a stochastic process. In general one can define $$\sigma(t)\doteq f(Y(t))\quad,$$
where $Y(t)$ is a generic driving process.
Given for the underlying price $S(t)$ the well-known Geometric Brownian Motion dynamics, it is convenient to make 
use of the zero-mean log-return $X(t)$:
\begin{equation}\label{X}
X(t) \doteq \ln\left[\frac{S(t)}{S_{0}}\right] - \mu t +\frac{1}{2}\int_{t_{0}}^{t}\sigma^{2}(t')dt'\quad ,
\end{equation}
whose Stochastic Differential Equation (SDE) reads
\begin{equation}\label{dX}
dX(t) = \sigma(t)dW_{1}(t)\quad .
\end{equation}
Therefore, the 2-dimensional volatility-return process can be written as
\beq\label{eq:sistema}
\left\lbrace \begin{aligned}
dX(t) &= f(Y(t))\, dW_{1}(t)\quad,\quad X(0)=0\\
dY(t) &= \alpha (m - Y(t) )\,dt + g (Y(t))\,dW_{2}(t)\quad,\quad Y(0)=Y_{0}\\
\end{aligned}\right .\quad. 
\eeq
Namely, $Y$ is taken as a mean reverting process, i.e. the deterministic term on the r.h.s of the 
second equation given in (\ref{eq:sistema}) 
is responsible for reverting the expectation value of $Y$ to the asymptotic value $m$ with relaxation time $1/\alpha$. 
The mean-reverting character of the SV models reflects the economic idea of a ``normal level'' of volatility, towards 
which an efficient market in healthy conditions tends.

The easiest way to incorporate in the models the correlation between volatility and returns, that is the familiar 
leverage effect, is to postulate that the two Wiener processes $W_{1,2}$ are correlated by a coefficient $\rho$ 
\beq\label{eq:corr}
dW_{2}(t) = \rho dW_{1}(t) + \sqrt{1 - \rho^{2}} dZ(t)\quad,
\eeq
where $Z(t)$ is a Wiener process independent of $W_{1}$. For this reason, SV models are given the attribute `correlated'. 
Eq.~(\ref{eq:corr}) is obtained recalling that each Wiener process must satisfy $\langle dW^{2}\rangle = dt$.

At present several SV models are discussed in the literature, differing in the dynamics attached to $\sigma$, namely in 
the choice operated for $f$ and $g$ in Eq.~(\ref{eq:sistema}). In this paper, we decided to take into consideration the Vasicek, 
Heston and exponential Ornstein-Uhlenbeck (exp-OU) models, three of the most popular within the family of SV models. They are 
listed in Tab.~\ref{tab:models}. 
\begin{table}
  \centering
  \caption{Models of volatility.}\label{tab:models}
  \begin{center}
    \begin{tabular}{l|lll}
      \hline
      Authors & $f(Y)$ & $Y$ process & $Y$ Stochastic Differential Equation\\
      \hline
      Vasicek & $Y$ & Mean-reverting OU & $dY(t) = \alpha (m - Y(t) )\,dt + k \,dW_{2}$\\
      Heston  & $\sqrt{Y}$ & CIR & $dY(t) = \alpha (m - Y(t) )\,dt + k \,\sqrt{Y} \,dW_{2}$\\
      exp-OU  & $e^{Y}$ & Mean-reverting OU & $dY(t) = \alpha \,Y(t) \,dt + k \,dW_{2}$\\
      \hline
    \end{tabular}
  \end{center}
\end{table}
It's worth mentioning that for each of them the volatility distribution can be obtained analytically, since its SDE 
features a single Wiener process. The same does not hold for the return distribution, which has to be worked out 
starting from the Fokker-Planck equation for the two-dimensional volatility-return process of Eq.~(\ref{eq:sistema}).

In Tab.~\ref{tab:theo_charac} we summarize the basic theoretical characteristics of each model, as obtained 
in \cite{perello,dragulescu,masoliver}. The main feature in the analysis of SV models is that the predicted 
log-return probability distribution functions (PDF) can be expressed only by means of their characteristic 
functions $\varphi_{X}(\omega,t)$, defined as
\begin{equation}\label{OUcharactdef}
\begin{aligned}
\varphi_X(\omega,t) &= \int d\sigma_0\,\varphi_X(\omega,t|\sigma_0) p_{\rm st}(\sigma_0)\\ &= \int d\sigma_0 \int dx \; e^{i\omega x} p_X(x,t|\sigma_0) p_{\rm st}(\sigma_0) \quad . 
\end{aligned}
\end{equation}
The first integral is obtained under the important hypothesis of considering the initial/final volatility at a 
normal level. Among the models considered, Eq.~(\ref{OUcharactdef}) can be inverted analytically only in the case of 
the exp-OU model, leading to a closed-form expression for the return distribution~\cite{masoliver}. 

Moreover, a good SV model must account not only for the volatility and return PDFs, but also must give realistic 
predictions for leverage effect and volatility autocorrelation observed in financial markets. These can be attained 
by defining the respective statistical coefficients
\begin{equation}\label{protolev}
\mathcal{L}(\tau)\doteq\frac{E\left[\sigma(t+\tau)^2dX(t)\right]}{E\left[\sigma(t)^2\right]^{2}}\simeq\frac{E\left[dX(t+\tau)^2dX(t)\right]}{E\left[dX(t)^2\right]^{2}}\quad,
\end{equation}
as in Ref.~\cite{bouchPRL}, and the analogous
\begin{equation}\label{volautocorr}
\begin{aligned}
\mathcal{C}(\tau)&\doteq\frac{\langle \sigma(t)^2\sigma(t+\tau)^2\rangle-\langle \sigma(t)^2\rangle^2}{
\mathrm{Var}\left[\sigma(t)^{2}\right]}\\
&\simeq \frac{\langle dX(t)^2dX(t+\tau)^2\rangle-\langle dX(t)^2\rangle^2}{
\langle dX(t)^4\rangle-\langle dX(t)^2\rangle^2}\quad .
\end{aligned}
\end{equation}
The correlations at different times between the Wiener processes allow to compute the expressions appearing in 
Tab.~\ref{tab:theo_charac}, where $\beta= k^{2}/2\alpha$. It is worth mentioning that in the case of the leverage 
coefficient, $\mathcal{L}(\tau)$ is null for $\tau<0$ ($H$ is the Heaviside step function), thus respecting the 
observed facts. As for the volatility autocorrelation it must be noticed that the exp-OU model yields, in a wholly 
natural way, two main time constants, thus resulting the most realistic of the three.
\begin{table}
  \centering
    \caption{Theoretical features of SV models.}\label{tab:theo_charac}
  \begin{center}
    \begin{tabular}{@{}c|cccc@{}}
      \hline
      & Vasicek & Heston & exp-OU\\
      \hline
      Volatility PDF\rule[-10pt]{0pt}{20pt} & Normal & Gamma & Log-normal\\
      Log-return PDF\rule[-15pt]{0pt}{30pt} & \parbox{.2\textwidth}{\centering $\varphi_{X}(\omega,t)$\\non-inv} & \parbox{.2\textwidth}{\centering $\varphi_{X}(\omega,t)$\\non-inv} & \parbox{.2\textwidth}{\centering $\varphi_{X}(\omega,t)$\\ invertible }\\
      $\mathcal{L(\tau)}$ \rule[-10pt]{0pt}{20pt} & $\rho e^{-\alpha \tau} \,H(\tau)$ & $?$ & $\rho e^{-k^{2} \tau} \,H(\tau)$\\
      $\mathcal{C(\tau)}$ \rule[-15pt]{0pt}{35pt}& \parbox{.2\textwidth}{\centering $\sim e^{-\alpha \tau}$\\ 1 time scale} & \parbox{.2\textwidth}{\centering $e^{-\alpha \tau}$\\ 1 time scale} & \parbox{.2\textwidth}{\centering {\large$\frac {\exp[4\beta \,e^{-\alpha \tau}] - 1}{3e^{4\beta}-1}$}\\  \rule[-5pt]{0pt}{15pt}2 time scales}\\
      \hline
    \end{tabular}
  \end{center}
\end{table}
%

\section{Numerical results}
\label{s:numerical}

The theoretical results were tested {\it ab initio} by means of original numerical simulations of the models, 
whose SDEs were discretized following a standard Euler-Maruyama method. As a rule of thumb, to have a sufficiently 
accurate simulation of the return-volatility paths the time step $\Delta t$ must be significantly shorter than the 
mean-reversion time $1/\alpha$. The model parameters used in the simulated SDEs were taken to assume values comparable 
with those appearing in the literature.

\begin{figure}[]
 \begin{minipage}[c]{.44\textwidth}
\centering
  \includegraphics [scale = 0.33]{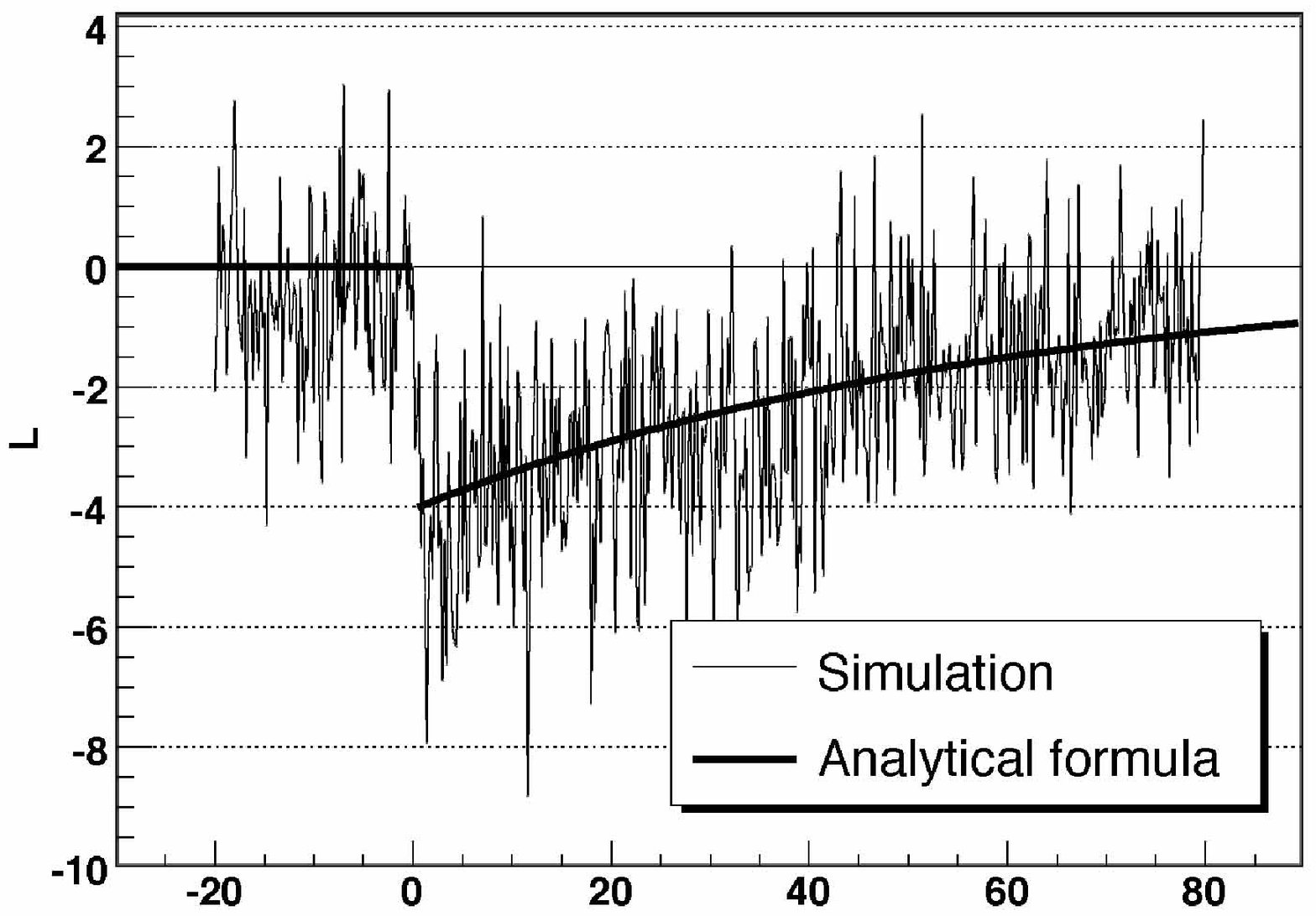}
 \end{minipage}
 \begin{minipage}[c]{.56\textwidth}
 \centering
 \includegraphics [scale = 0.35]{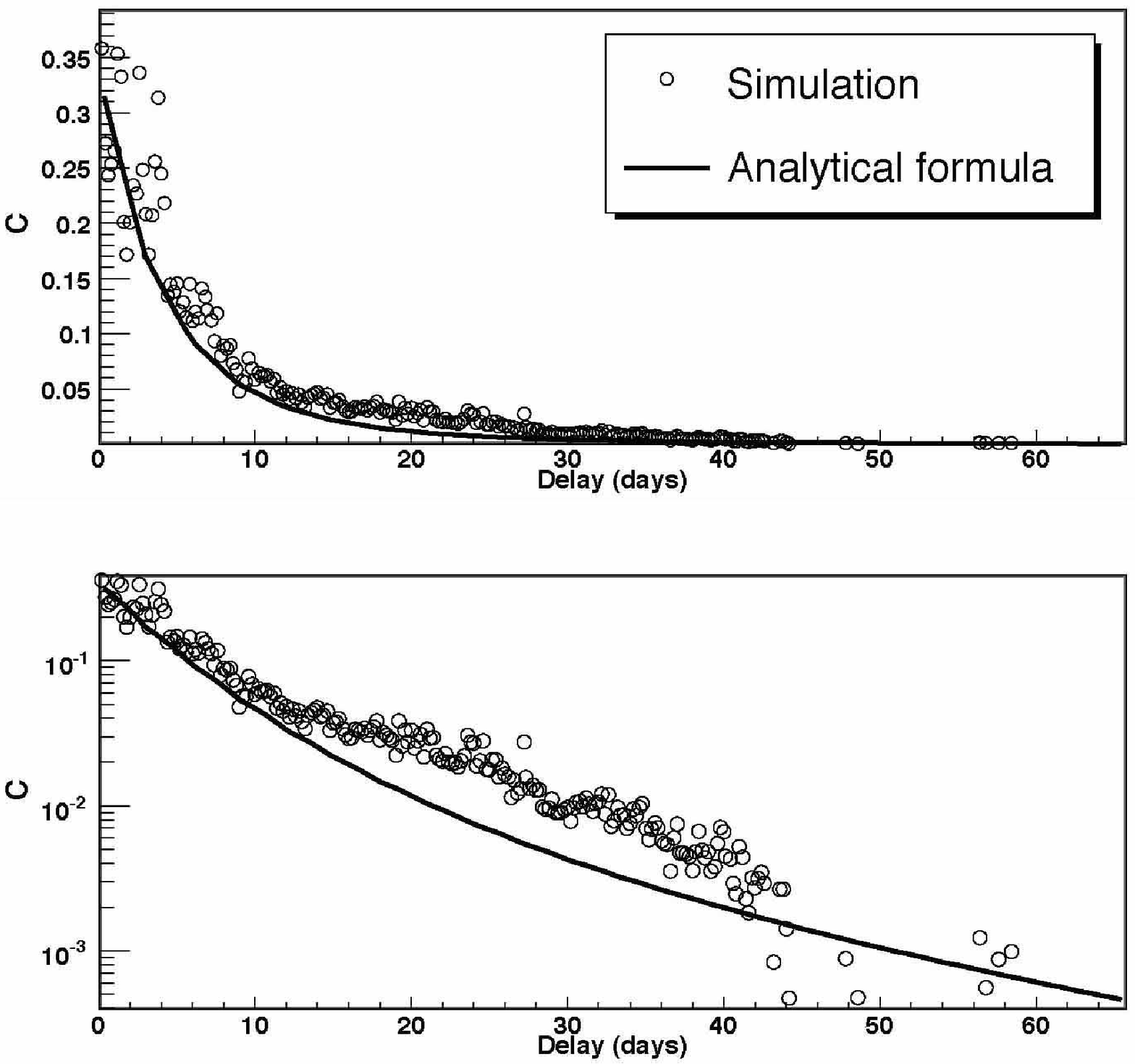}
 \end{minipage}
\caption{\label{fig:corr}Comparison between simulated different-time correlations for exp-OU model and their analytical forms. Left panel: leverage effect analytical expression compared with the numerical leverage function. Right panel: volatility autocorrelation function compared with the corresponding analytical form.}
\end{figure}

The produced paths were used to generate Monte Carlo populations whose distributions (or distribution parameters) were 
graphically compared with the analytical results regarding the volatility and log-return processes. We tested the time 
evolution of the mean and variance values of the volatility distributions, the shape of the same distributions for 
several time instants, and the log-return distributions at different times. The simulated paths were evolved over time 
intervals ranging from a few days to about one financial year. The most delicate step in the procedure was indeed the 
mere graphical contrast, since it requested to invert the analytical characteristic functions of log-return distributions 
in the Vasicek and Heston models. Such operation was done with the help of the built-in Fast Fourier Transform functions 
of \ROOT and \MAT. The shape of the leverage and volatility autocorrelation coefficients versus the time delay $\tau$ 
(see Eqs. \ref{protolev} and \ref{volautocorr}) was also obtained. The expectation values appearing in the correlation 
functions were calculated on a single, very long simulated return series, corresponding to $\sim$100 years of trading, 
as widely used in the financial practice. The entire analysis and more technical details can be found in Ref.~\cite{lorenzo}.
\begin{figure}[h!]
 \begin{minipage}[c]{.48\textwidth}
 \centering
 \includegraphics[scale = 0.34]{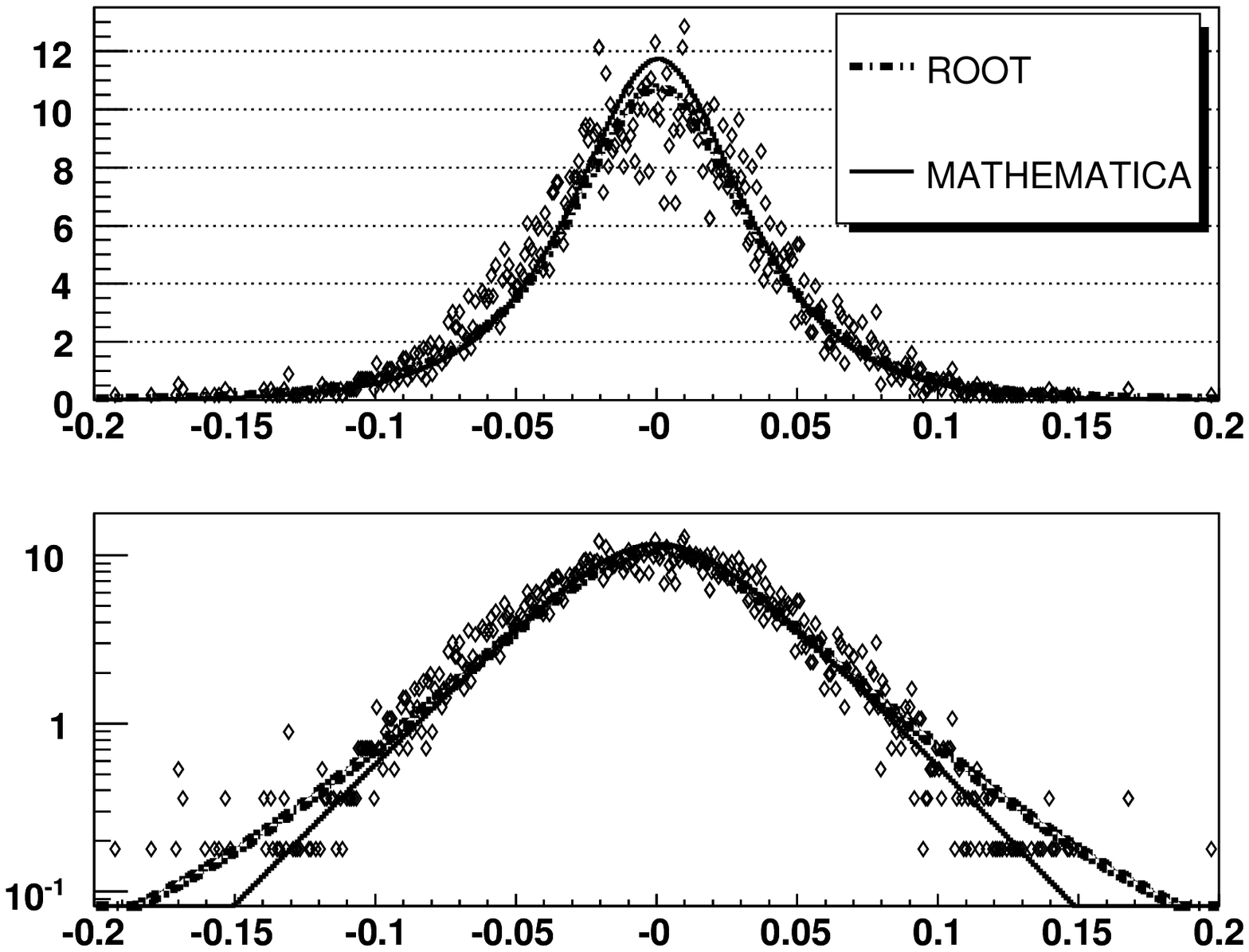}
 \end{minipage}
 \begin{minipage}[c]{.48\textwidth}
 \centering
 \includegraphics[scale = 0.34]{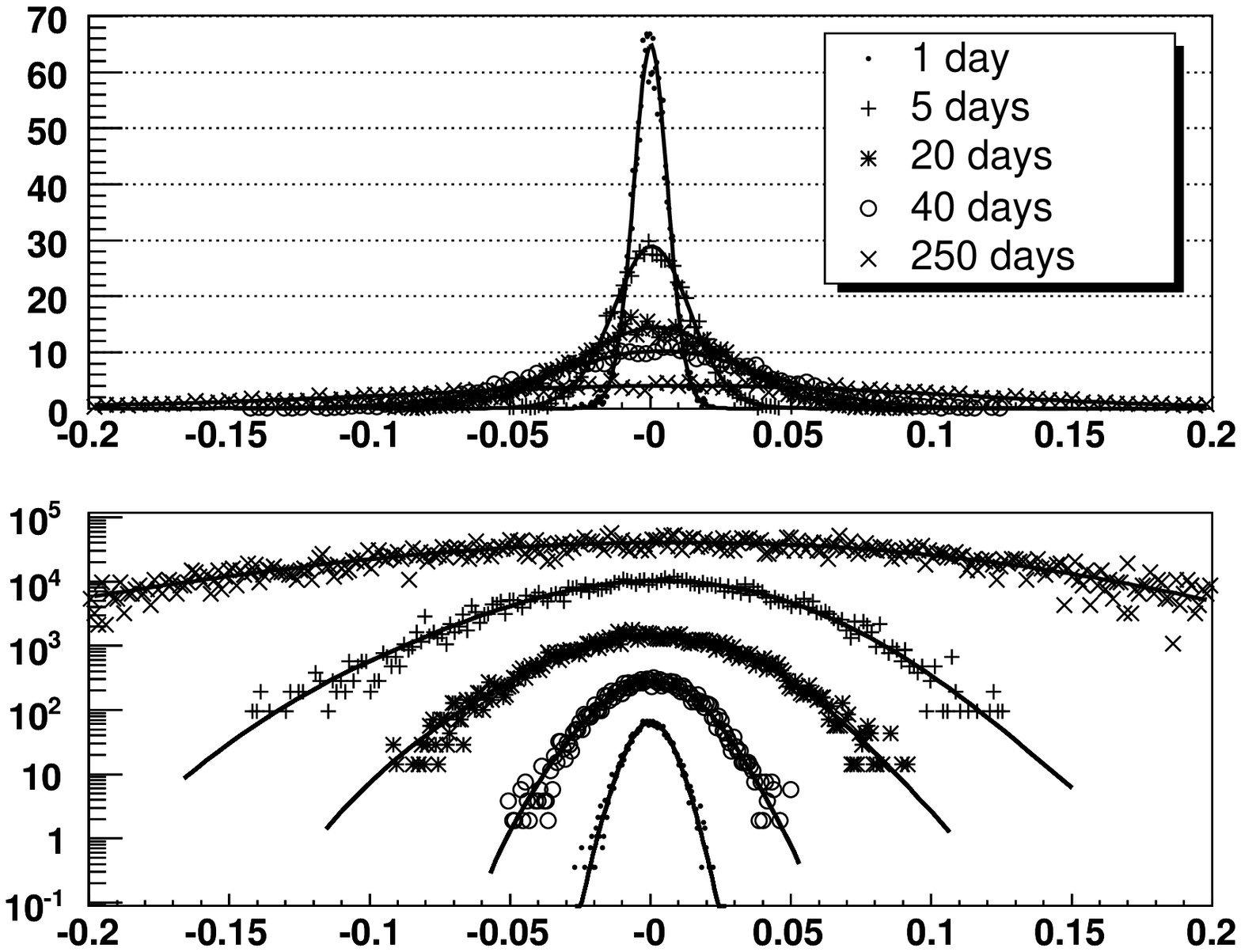}
\end{minipage}
 \caption{\label{fig:pdfs}Comparison between simulated return PDFs (dots) and theoretical results (lines). Left panel: Heston model simulated on a 20-day time horizon. Right panel: exp-OU model simulated at various time horizons vs the analytical expression of~\cite{masoliver}}
 \end{figure}
\newline
In Fig.~\ref{fig:pdfs}-\ref{fig:corr} we show some examples of the analysis carried out: all the numerical outcomes 
agree very well with the predictions of Tab.~\ref{tab:theo_charac} for each model, showing in the same time that these 
are correct and, conversely, that the models can be effectively simulated even with a quite simple strategy. These 
conclusions hold for both log-return PDFs (Fig.~\ref{fig:pdfs}) and the different-time correlation functions between 
log-return and volatility, i.e. the leverage effect, and between volatility and itself (Fig.~\ref{fig:corr} left and right, 
respectively).

\section{Empirical analysis}
\label{s:empirical}

We also performed an empirical analysis of financial data in order to critically compare the models and to 
establish whether they can be successful in predicting the stylized facts of real markets.
The time series  used, freely downloaded from Yahoo Web Site\footnote{http://finance.yahoo.com/}, are collections of daily 
closing prices of the Italian assets Bulgari SpA, Brembo and Fiat SpA from January 2000 to May 2007. Here, we present in 
detail the analysis of Fiat SpA data; the entire study performed on Italian shares can be found in Ref.~\cite{enrica}.

\begin{figure}[h!]
  \centering
  \includegraphics [scale = 0.26]{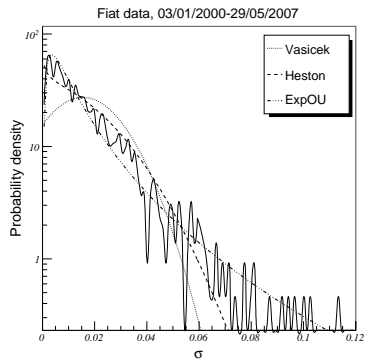}
  \caption{\label{fig:fit_vol}Fit~to~empirical~daily~volatility~PDF~of~Fiat~SpA.}
 \end{figure}

Fig.~\ref{fig:fit_vol} shows a comparison between the theoretical distributions displayed in 
Tab.~\ref{tab:theo_charac} and the empirical daily volatility for Fiat SpA, evaluated by means of the proxy described in 
Ref.~\cite{bouchaud} as absolute daily returns. It's worth mentioning that a similar analysis is also performed 
in Ref.~\cite{micciche}. The parameter values of the fitted curves are obtained according to a multidimensional 
minimization procedure based on the maximum likelihood approach. From Fig.~\ref{fig:fit_vol} the best agreement 
between empirical data and the theory clearly emerges for the Log-Normal distribution predicted in the exp-OU 
framework, whereas the Normal and the Gamma densities tend to underestimate the large distribution's tail, as 
already remarked in Ref.~\cite{bouchaud}. 

\begin{figure}[ht!]
  \begin{minipage}[b]{0.5\textwidth}
  \begin{center}
    \includegraphics[scale = 0.33]{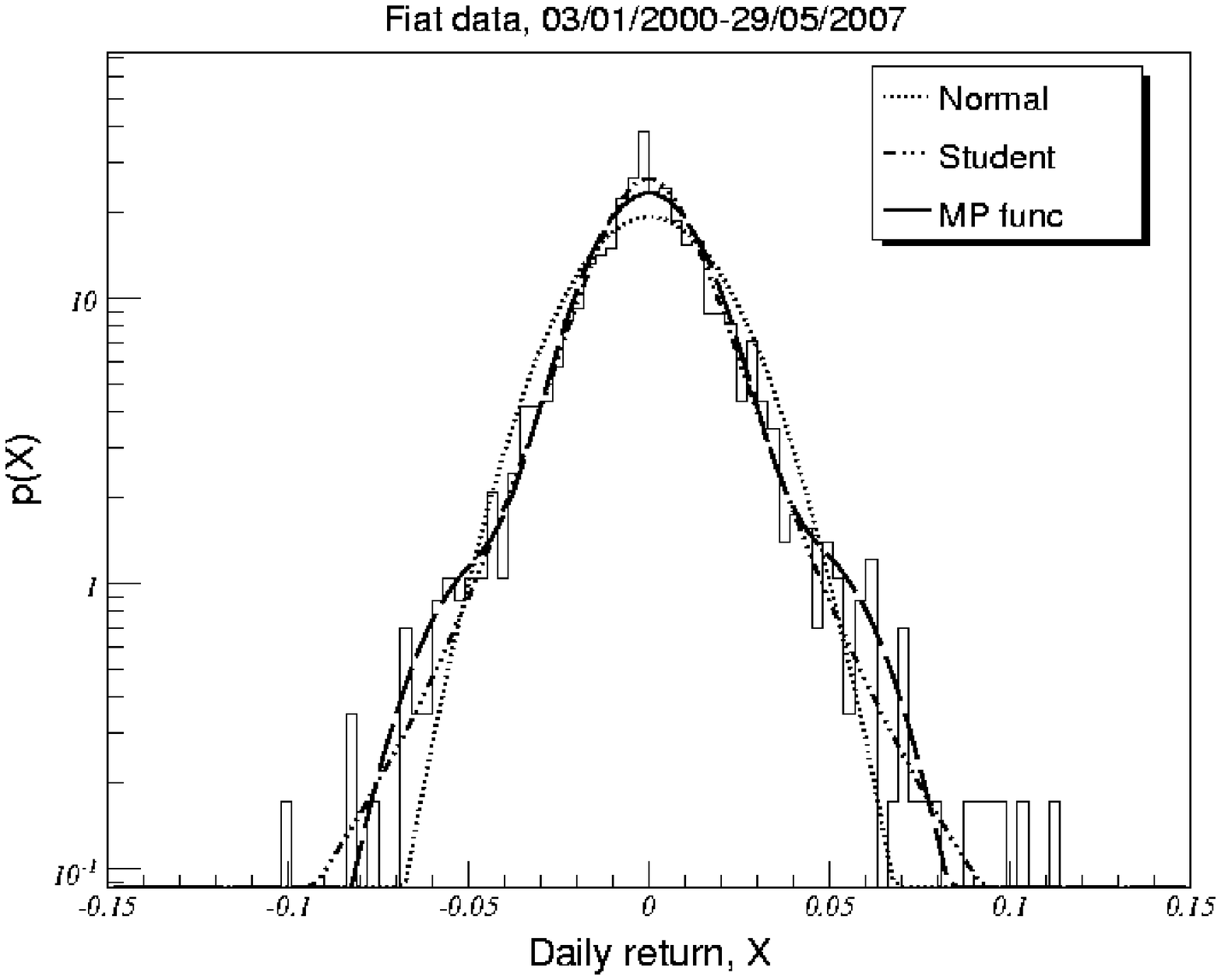}
    \end{center}
    \end{minipage}
  \begin{minipage}[b]{0.5\textwidth}
    \begin{center}
      \includegraphics[scale = 0.33]{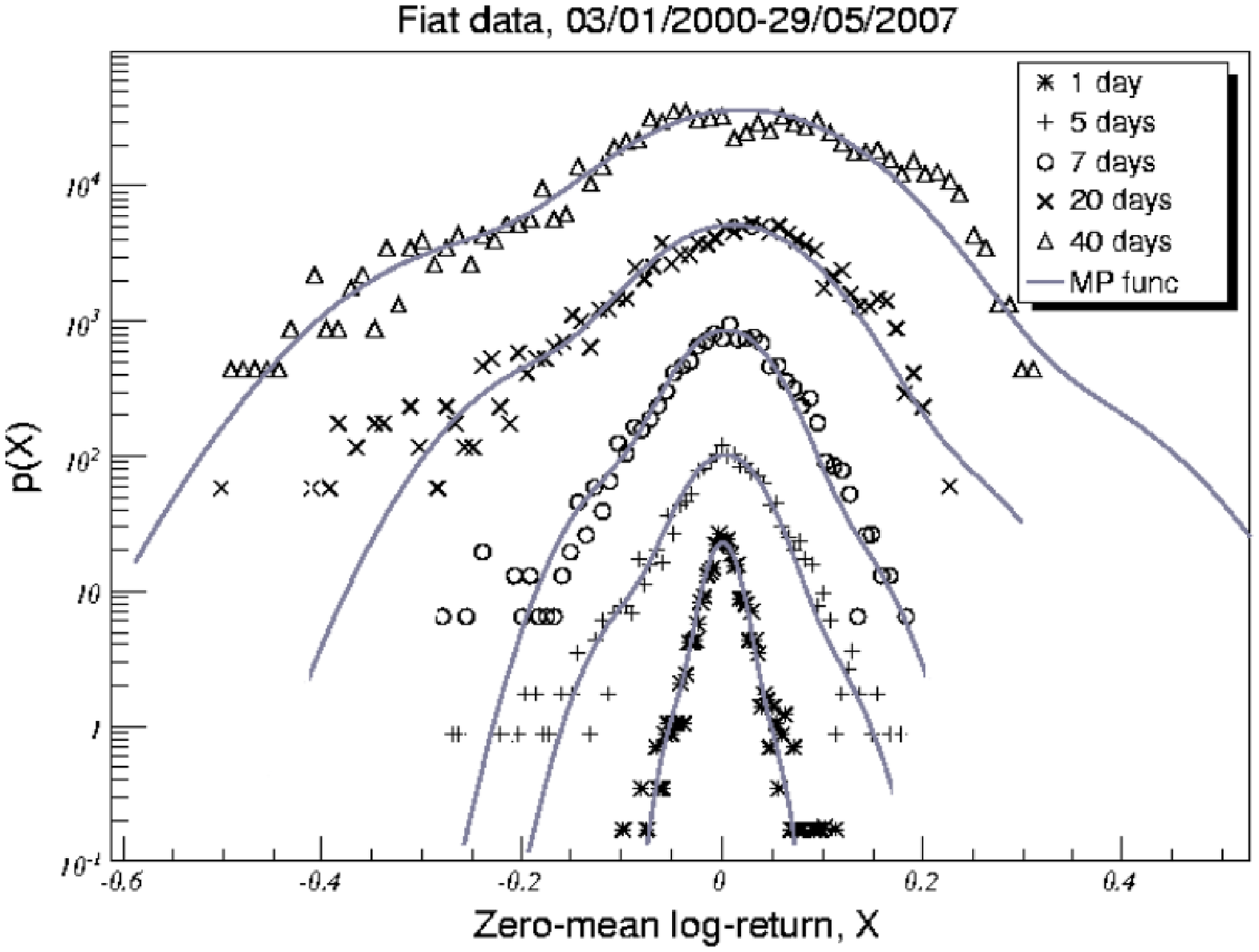}
      \end{center}
    \end{minipage}
    \caption{\label{fig:multifit_ret}Comparison between empirical returns and the MP theory~\cite{masoliver} for Fiat SpA. Left panel: fit of daily returns (histogram) on MP curve in comparison with the Normal and the Student-$t$ PDFs. Right panel: returns (dots) for different time horizons, shifted each other by one decade, vs the MP prediction (line).}
  \end{figure}

In the light of this result we propose a comparison between historical log-return probability density 
and the analytical formula derived by Masoliver and Perell\'o (MP) in the exp-OU framework~\cite{masoliver}. 
Figure~\ref{fig:multifit_ret} (left panel) compares daily log-returns with the MP theory, 
together with the Normal and the fat-tailed Student-$t$ PDFs~\cite{mantegna,giacomo}. To solve the optimization 
problem and find the best parameter values required for the fit, we implement a numerical algorithm based on the 
MINUIT program of the CERN library. In particular, for the MP function we perform a multidimensional fit over four 
free parameters, finding out values in good agreement with those quoted in the literature~\cite{masoliver}. From 
Fig.~\ref{fig:multifit_ret} (left panel) it emerges that the Student-$t$ and MP curves are in good agreement 
with the empirical return distribution for both the central body and the tails of the histogram, while the Normal 
distribution fails to reproduce the data. Moreover, it's also quite evident that Student-$t$ better captures the 
extreme events of the distribution, due to its strong leptokurtic nature.

In the right panel of Fig.~\ref{fig:multifit_ret} we show the log-return distributions for several time horizons (points) 
in comparison with the MP theory (solid line). It's worth mentioning that all the theoretical curves appearing in the 
figure aregenerated by changing in the analytical formula only the value of the temporal parameter according to the 
time lag under analysis, while for the other parameters we use the same values evaluated from the fit of daily data. 
In this way, we can directly compare empirical data and theoretical predictions: the overall agreement
is very good. The success of the MP theory emerges also noting that when the time lag increases, the left tail 
of the empirical distributions becomes fatter and the analytical solution tends to increase the absolute value of 
its skewness, becoming more asymmetric and similar to the data shape. 

\section{Conclusions}
\label{s:conclusion}

For all of the considered SV models we proved an almost perfect convergence between theoretical results and numerical 
simulations, of particular interest in view of financial applications like risk management and option pricing. Among them, 
exp-OU has been found to be the most successful in predicting the empirical volatility 
distribution. The theoretical analysis of the model yields also a double time scale in the 
volatility autocorrelation, a well-known fact in real financial data. The model also fits the log-return distribution 
quite well, even if a better agreement with the PDF tails for infra-week data would require to 
substitute the Wiener noise in the model SDEs with a non-Gaussian one, therefore leaving the Black-Scholes-Merton framework.

\end{document}